\documentstyle[prl,aps]{revtex}
\begin{document}
\draft
\title{
Importance of the Doppler Effect to the Determination of the Deuteron Binding
Energy}
\author{Yongkyu Ko \footnote{e.mail : yongkyu@phya.yonsei.ac.kr}, 
Myung Ki Cheoun and Il-Tong Cheon}
\address{
Department of Physics, Yonsei University, Seoul 120-749, Korea}
\date{\today}
\maketitle
\begin{abstract}
The deuteron binding energy extracted from the reaction ${}^1H(n,\gamma){}^2H$ 
is reviewed with the exact relativistic formula, where the initial kinetic
energy and the Doppler effect are taken into account.  We find that the
negligible initial kinetic energy of the neutron could cause a significant 
uncertainty which is beyond the errors available up to now.
Therefore, we suggest an experiment which should include the detailed 
informations about the initial kinetic energy and the detection angle.  
It could reduce discrepancies among the recently reported values about the 
deuteron binding energy and pin down the uncertainty
due to the Doppler broadening of $\gamma$ ray.
\end{abstract}
\vspace{1cm}
\pacs{ PACS numbers: 21.10.Dr, 11.80.Cr, 25.40.Lw, 27.10.+h }
The deuteron binding energy is one of the most important physical quantities
in nuclear physics.  This plays a central role in the determination of
the neutron mass, and provides a critical reference energy for the
determination of high-energy $\gamma$ rays measured on the atomic-mass scale
\cite{Greene}.  There are several articles to have measured and improved
the deuteron binding energy  more precisely than
ever \cite{Greene,Van,Vylov2,Vylov1,Greenwood,Adam,Michael}.
Some of them have reported their values as precisely as up to the order of 
several electron volts \cite{Greene,Van,Vylov1,Adam}.  Usually, the deuteron 
binding energy is determined by adding the recoil energy of the deuteron
to the measured energy of the $\gamma$ ray which is emitted from the neutron
capture reaction, ${}^1H(n,\gamma){}^2H$.  In the conventional non-relativistic 
treatment, there seems to be nothing to correct any more, because the initial 
kinetic energy of the neutron is assumed to be so small that it can be 
neglected safely.  Actually, the kinetic energy of the neutron used in 
Ref. \cite{Greene} is 0.056 eV and that of the thermalized
neutron from the neutron source in Ref. \cite{Van,Vylov1,Greenwood} is assumed 
to be 0.025 eV at room temperature.

However, if the two body collision process is treated relativistically,
we demonstrate that such a small kinetic energy of the neutron
causes a considerable uncertainty in the deuteron binding energy because of
 the Doppler effect. Even the very small neutron kinetic energy gives rise to
a significant velocity of the n-p system in the laboratory frame.  As a result, 
this moving source of the $\gamma$ ray can be an origin for the Doppler effect. 
Naturally this Doppler effect depends on the angle of the $\gamma$ ray detector 
with respect to the velocity of moving source.  Therefore, the measured 
$\gamma$ ray energy is expected to have an unavoidable uncertainty originating 
from this angle dependence.  
As will be shown later on, our estimation for the  uncertainty shows about 
25 eV maximally for the neutron initial kinetic energy of 0.056 eV and 14.0 eV
for that of 0.025 eV, respectively. These values are so large compared to the 
reported error 2.3 eV \cite{Greene}, which comes from the $\gamma$ ray detector 
itself. 

Let us suppose that two particles are initially at rest within an
appropriate interaction distance, and they come together to form one
body system by emitting a photon.
In this situation, the center of momentum frame \cite{Goldstein}, which is an
exact nomenclature in relativistic kinematics but used as the center of
mass frame hereafter, coincides with the laboratory frame.  From the energy
and momentum conservation, we get the following relation
\begin{equation}
p_1 + p_2 = q + k \label{va},
\end{equation}
where $p_i$'s are four momenta of the initial two particles while $q$ and $k$
are those of the resultant composite particle and the photon, respectively.
The time component of the momentum accounts for the energy conservation, and
its space components stand for the momentum conservation.
The square of Eq. (\ref{va}) is Lorentz invariant:
\begin{equation}
(p_1 + p_2)^2 = (q + k)^2\label{ka}.
\end{equation}
In the case of the initial particles at rest, this gives the following equation 
in the center of mass frame
\begin{equation}
(m_1 + m_2)^2 = M^2 + 2 \omega^2 + 2 \omega \sqrt{M^2 + \omega^2},\label{da}
\end{equation}
where $M$ and $m_i$ are masses of the composite and the initial particles,
respectively, and $\omega$ is the energy of the emitted photon.
Solving for $\omega$, we obtain the exact equation for the
energy of the emitted photon in terms of the masses of the initial and the final
particles:
\begin{eqnarray}
\omega &=& {(m_1 + m_2 + M)(m_1 + m_2 - M) \over{2(m_1 + m_2)}}\nonumber\\
       &=& {(E_b + 2 M)E_b \over{2(E_b + M)}}\label{ha},
\end{eqnarray}
where the binding energy $E_b$ is defined as the mass difference between the 
initial particles and the final product $E_b = m_1 + m_2 - M$.
From Eq. (\ref{ha}), we get the binding energy in terms of the produced
mass and the energy of the emitted photon:
\begin{eqnarray}
E_b &=& \omega + \sqrt{\omega^2 + M^2} - M \nonumber\\
    &\cong& \omega + {\omega^2 \over{2M}},\label{ga}
\end{eqnarray}
where the last expression shows the non-relativistic approximation which
is used to determine the binding energy of the deuteron so far 
\cite{Greene,Van,Vylov2,Vylov1,Greenwood,Adam,Michael}. 

In the case that the incident particle has a kinetic energy and comes to
the target at rest, the laboratory frame differs from the center of mass
frame.  Using the Lorentz invariance of the square of four momentum,
we obtain the following equation
\begin{equation}
(p_1 + p_2)^2_{lab} = (p_1 + p_2)^2_{c.m} = (q + k)^2_{c.m}, \nonumber\\
\end{equation}
which is calculated as
\begin{equation}
m_1^2 + m_2^2 + 2 m_2E_{1l}
= M^2 + 2 \omega^2 + 2 \omega \sqrt{M^2 + \omega^2}.
\end{equation}
Notice that $\omega$ is the energy of the photon emitted in the center of mass
frame and is given by
\begin{equation}
\omega = {m_1^2 + m_2^2 - M^2 + 2 m_2E_{1l} \over{2\sqrt{m_1^2 + m_2^2
+ 2 m_2E_{1l}}}}.
\end{equation}
Here the velocity of the center of mass frame is calculated as
\begin{equation}
v_{c.m} = {|\mbox{\boldmath{$p$}}_{1l}| \over{m_2 + E_{1l}}}.
\end{equation}
The above energy of the photon should be transformed to that of the laboratory
frame in which the observation is carried out.
This transformation is nothing but the Doppler effect and is given, in
relativistic theory, by
\begin{equation}
\omega_l = \omega {  \sqrt{1 - v_{c.m}^2} \over{1 - \mbox{\boldmath{$n$}}
\cdot \mbox{\boldmath{$v$}}_{c.m}}},
\end{equation}
where $\mbox{\boldmath{$n$}}$ is the unit vector in the direction of the
photon beam.  Thus, we obtain the energy of the photon measured in the
laboratory frame:
\begin{eqnarray}
\omega_l &=& \omega {\sqrt{1 - v_{c.m}^2}
           \over{1 - v_{c.m}\cos \theta}}\nonumber\\
       &=& {m_1^2 + m_2^2 - M^2 + 2 m_2E_{1l}
           \over{2(m_2 + E_{1l} - |\mbox{\boldmath{$p$}}_{1l}|\cos\theta)}}
           \nonumber\\
       &=& {(E_b + 2 M)E_b + 2 m_2 K_1 \over{2(E_b + M + K_1 -
       |\mbox{\boldmath{$p$}}_{1l}|\cos \theta)}},\label{sa}
\end{eqnarray}
where the kinetic energy is defined by $K_1=E_1-m_1$.
It should be noted that $\omega_l$ depends on $\theta$, if the neutron beam 
direction is fixed.  Alternatively, the same expression can be derived 
directly from Eq. (\ref{ka}) in the laboratory frame.  We consider here 
three special cases for easy understanding as explained in Ref. \cite{Beiser}.  
The first case, $\theta = {\pi \over 2}$, which represents the transverse 
Doppler effect, is when we measure the frequency of the photon emitting 
perpendicular to the incident beam.
The second case, $\theta = \pi$, which corresponds to the longitudinal Doppler
effect in which the source is receding, is when we measure the photon emitting
anti-parallel to the incident beam.  The last case, $\theta = 0$,
which is also longitudinal but the source is approaching, is when we measure
the photon emitting parallel to the incident beam.  Solving for the binding
energy, we obtain
the following equation for the binding energy without any approximation
\begin{equation}
E_b = \omega_l - M + \sqrt{M^2 + \omega_l^2 -2(m_2 - \omega_l)K_1 - 2 \omega_l
      |\mbox{\boldmath{$p$}}_{1l}| \cos \theta}, \label{na}
\end{equation}
where $|\mbox{\boldmath{$p$}}_{1l}| = \sqrt{K_1(K_1 + 2m_1)}$.
This equation, which can be reduced to Eq. (\ref{ga}) as the initial kinetic
energy $K_1$ goes to zero, contains the initial kinetic energy of the neutrons
and the detection angle.  Both of them were not considered in the 
non-relativistic analysis of the experiment of $ {}^1H(n,\gamma){}^2H $ 
reaction \cite{Greene}.

In the following, we show the detailed numbers coming from the above $\theta$ 
dependence and compare to the reported error.
The measured wavelength of the $\gamma$ ray emitted from the n-p capture has 
been reported as \cite{Greene}
\begin{equation}
\lambda_{np} = 5.576~698~8~~(55) \times 10^{-13} m.
\end{equation}
The error in parenthesis comes from the Bragg angle and the lattice-spacing 
of crystal which are related with the equation $n \lambda = 2 d 
\sin \theta_{Bragg}$.  Using the conversion constant $\hbar c$, we obtain the 
energy of the photon:
\begin{equation}
\omega = {2 \pi \hbar c \over{\lambda_{np}}}=2.223~255~2~(23) \rm{MeV}.
\end{equation}
The physical constants used in this letter are \cite{prd}
\begin{eqnarray}
& & \hbar c = 197.327~053~(59) \rm{MeV~fm},   \nonumber\\
& & \text{deuteron mass}~M = 1875.613~39~(57) \text{MeV}, \nonumber\\
& & \text{neutron mass}~m_1 = 939.565~63~(28) \text{MeV}, \nonumber\\
& & \text{proton mass}~m_2 = 938.272~31~(28) \text{MeV}.
\end{eqnarray}
Since a deuteron mass $M$ in Eq. (\ref{na}) causes an uncertainty in summing 
the significant figures, we expanded the equation in order to visualize
the cancellation of the uncertainty due to the deuteron mass as follows
\begin{eqnarray}
E_b &\cong& \omega_l + {\omega_l^2 \over{2 M}} - {(m_2 - \omega_l)K_1 \over M}
      - {\omega_l \sqrt{K_1(K_1 + 2m_1)} \over M} \cos \theta  \nonumber\\
&=& 2.223~255~2~(23) + 0.001~317~7 -  0.000~000~03 - 0.000~012~2 \cos \theta
 ~ \text{MeV} \label{ba}.
\end{eqnarray}
In this result, the second term is the deuteron recoil effect, the third term
comes from the initial kinetic energy itself and the fourth term stands for the
Doppler effect due to the moving source of the center of mass frame.
Since the authors
of Ref. \cite{Greene} have used the neutron flux of which distribution is
approximately Maxwellian with a peak at $1.2$ \AA (= 0.056 eV), we calculate
the binding energies for the three cases of $ \theta = 0,
{\pi \over2},\pi $:
\begin{eqnarray}
& &E_b = 2.224~560~1~(23)~ \rm{MeV}~\text{for}~\cos \theta = 1,\nonumber\\
& &E_b = 2.224~573~2~(23)~ \rm{MeV}~\text{for}~\cos \theta = 0,\nonumber\\
& &E_b = 2.224~585~2~(23)~ \rm{MeV}~\text{for}~\cos \theta = - 1. \label{dop}
\end{eqnarray}
This Doppler effect will be reflected as a line width, a line shift of $\gamma$ 
ray or both of them in the experiments.
As we know from the above numerical calculations, the Doppler effect due to the
initial kinetic energy of the neutron causes an uncertainty by around 25 eV,
which is completely out of the range of the above error 2.3 eV.  This error may 
be the best value of the present measurement technique of the $\gamma$ ray in 
that energy scale.   Consequently,
it is inescapable to take the uncertainty due to this Doppler effects into 
account in the analysis of the ${}^1H(n,\gamma)^2H$ experiments.  

In the actual experiment using the thermalized neutron, however, the dependence 
of the angle is not detected, but manifests itself as a line 
width in $\gamma$ ray spectrum because of the following reason.  Since the 
incident neutrons may be distributed at random and isotropic in this 
experiment, we can assume that they are uniformly distributed over the whole 
solid angle $4 \pi$ with respect to the direction of the detected photons.  
Therefore the $\gamma$ ray detector fixed at the given angle sees all $\gamma$ 
rays triggered by the incoming neutrons in the whole angle range, i.e., from 
the angle $\theta=0$ to $\theta= \pi$ in Eq. (\ref{ba}).  Among those neutrons, 
the ones from the angle $\theta ={\pi \over 2}$ are most probable because of 
azimuthal symmetry, that is, the number of the incoming neutrons is 
proportional to $2 \pi \sin \theta d\theta$.  
Consequently, we estimate the width of the spectrum 
of the emitted photons from Eq. (\ref{sa}) as follows
\begin{eqnarray}
\Delta \omega_l &=& \sqrt{ \{ \omega'_l(K_0,\cos \theta_0) \}^2 \Delta K^2 +
   \{ \omega'_l(K_0,\cos \theta_0) \}^2 \Delta \cos \theta^2 }\nonumber\\
   &=& \sqrt{0.5^2 \Delta K^2 + 12.2^2 \Delta \cos \theta^2} ~\text{eV}
   \nonumber \\
   &=& 21.1 ~\text{eV},~~\text{for } \Delta \cos \theta= \sqrt{3}~~
   \text{and}~~ \Delta K = 0.005~\text{eV} \label{pa},
\end{eqnarray}
where we expanded it at $K_0 = 0.056$ eV and $\cos \theta_0 = 0$.  For the 
initial neutron kinetic energy of 0.025 eV, the line width is expected to be 
14.0 eV.  The above numbers are surprisingly large compared to the error from 
the crystal spectrometer.  Here one can expect another line width coming from 
the uncertainty of the kinetic energy of the incident neutrons, expressed by 
$\Delta K$ in Eq. (\ref{pa}).  But it is very small enough to be 
neglected as calculated above and pointed out in Ref. \cite{Greene}.  Thus, 
from the above relation, the
broadening of the $\gamma$ ray is mainly due to the Doppler effect.  However
most of Ge detectors or the other $\gamma$ ray detectors may not resolve the 
broadened spectrum, so the crystal 
spectrometer is used for precise measurements by determining the Bragg angle 
as in Ref. \cite{Greene}.  

In this experimental situation, the broadened 
$\gamma$ ray causes the uncertainties in the measurement of the Bragg angle 
and also in the spectrums of the $\gamma$ ray detector which determines the 
peak positions.  The authors of Ref. 
\cite{Greene} have measured the 52 Bragg 
angles between the two centers of the peaks of the broadened spectrums.  The 
small 
error of 2.3 eV, which comes from the statistical error in the data of the 52 
Bragg angles and the lattice-spacing of crystal, is much less 
than the calculated line width of 14.0 - 21.1 eV.  The reason of such a small 
error may be conjectured that the standard error is obtained from the standard 
deviation by dividing it by the square root of the number of measurements 
and the errors in the center positions of the broadened spectrums are  
neglected.  So the standard deviation, which is simply calculated as 
$2.3 \times \sqrt{52} = 16.6$ eV, is compatible with our calculated line 
width.  Therefore if we reduce the line width 
of the $\gamma$ ray by selecting the special angle $\theta={\pi \over 2}$, 
the measurement of the Bragg angle has much smaller error and is much more 
convincible though it is 
difficult to reduce the uncertainty of the $\gamma$ ray detector itself.  
Actually, the standard deviation of the measured Bragg angle $1.4 \times 
10^{-8}$ rad in Ref. \cite{Greene} is larger than the sensitivity and accuracy 
of $\sim 10^{-4}$ arcsec ($\leq 10^{-9}$ rad) of the Michelson interferometer,
so that it can be reduced up to the sensitivity.

Table \ref{table1} shows the update values of the deuteron binding energy.
Among them, only two articles \cite{Van,Vylov2} have shown their
experimental geometry.  So, we discuss these experimental geometry on 
a possible Doppler effect.  Since these experiments seems to be carried out 
at room temperature, we assume that the kinetic energy of the thermal 
neutrons is 0.025 eV because there are no more informations on the kinetic 
energy of the incident neutron.  In the experiment of Ref. \cite{Van}, the 
incident neutrons seem to be well-thermalized and the neutron source is 
located to the side of the paraffin sample.
The neutrons moving perpendicular to the direction of the detected photons are 
captured dominantly as the previous consideration, and the width of the 
detected photons is calculated as 
14.0 eV with using Eq. (\ref{pa}).  
If the paraffin sample is more thin and has 
larger diameter than that of this experiment and exposed to the thermalized 
neutrons moving perpendicular to the detected photon by shielding the sample 
except the round surface of the sample cylinder, one can reduce 
$\Delta \cos \theta$ and the Doppler broadening of the spectrum of 
the photon.  

In the experiment of Ref. \cite{Vylov2}, the situation is so 
complicated to treat incident neutrons as well-thermalized ones.
There is a possibility that many fast neutrons can be captured
through thermalization in the paraffin sample, because the source emits
much higher energy neutrons than the thermal neutrons.  Even though it is 
difficult to know what energy of the neutrons contribute to the capture process
dominantly, it is possible to compare this experimental result with others in 
the point of view of the Doppler shift.
In Refs. \cite{Van,Greenwood}, the authors have modified the binding energy 
of Ref. \cite{Vylov2} on the basis of the same standards of the calibration
energies of Refs. \cite{Helmer,Kessler} as 2.224 628 (15) MeV, and have 
compared it with their measured ones.  The corrected value 
is shifted by 53 eV with respect to the value of Ref. \cite{Van} and 64 eV with 
respect to the value of Ref. \cite{Greenwood}.  
These shifts occur, when the kinetic energy of the dominant incident neutron 
is more than at least 1.104 eV under the assumption that the incident
neutrons are parallel to the direction of the detected photons according to 
our result.
The authors of Ref. \cite{Vylov2} have also revised their value as 
2.224 568 (8) MeV on the basis of the standards of Refs. \cite{Helmer,Vylov3} 
in Ref. \cite{Vylov1}.  In the point of view of our result, 
this value can not be understood easily, on the contrary, this discrepancy 
is attributed to the problem of the calibration using a common system of 
standard energies.  Table \ref{table3} shows other similar experiments, 
in which it is more important to consider the Doppler effect 
in the case of a triton.

The direction of motion of the thermal neutrons is distributed at random
in the previous experiments 
\cite{Greene,Van,Vylov2,Vylov1,Greenwood,Adam,Michael}, and Doppler 
broadening contributes in
full to the observed line width.  Selecting a certain angle of the
incident neutron with respect to the emitted photon will lead to a
reduced line width and should therefore make it possible to obtain even
more precise results.  Such an experiment would be very challenging.  A
possible geometry could make use of a target sample near the core of a
reactor with appropriate shielding from thermal neutrons.  Using the
difference of the penetration lengths for neutrons and photons, shielding
the target should not cause much loss in flux of the incident neutron
beam.  As shown by the detailed numbers, it is indispensable to reduce the
Doppler broadening for a more precise measurement of the gamma-ray
wavelength and the determination of the deuteron binding energy.  Careful
geometric considerations can also explain the discrepancies among the
values of the deuteron binding energy reported in Refs. 
\cite{Greene,Van,Vylov2,Vylov1,Greenwood,Adam,Michael}, if
calibration problems are properly accounted for.

\begin {center}
\acknowledgments
\end{center}
We would like to thank Drs. B. S. Han, H. C. Kim and S. J. Hyun for the careful 
readings of the manuscript and discussions.  This work was supported by the 
Korean Ministry of Education(Project no. BSRI 97-2425).

\begin{table}
\begin{center}
\begin{tabular}{|c|c|}
\hspace{1cm} Reference \hspace{1cm} & Binding energy (eV)\hspace{3cm} \\ 
\hline\hline
\hspace{1cm} Greene {\it et al.} \cite{Greene}   \hspace{1cm}    & 2 224 589.0 (2.2)
\hspace{3cm} \\
\hspace{1cm} Van der Leun {\it et al.} \cite{Van} \hspace{1cm}   & 2 224 575 (9)  
\hspace{3cm}   \\
\hspace{1cm} Vylov {\it et al.} \cite{Vylov2}  \hspace{1cm}      & 2 224 572 (40) 
\hspace{3cm}   \\
\hspace{1cm} Vylov {\it et al.} \cite{Vylov1}   \hspace{1cm}     & 2 224 568 (8)  
\hspace{3cm}   \\
\hspace{1cm} Greenwood {\it et al.} \cite{Greenwood}\hspace{1cm} & 2 224 564 (17) 
\hspace{3cm}   \\
\hspace{1cm} Adam {\it et al} \cite{Adam}       \hspace{1cm}     & 2 224 574 (9)  
\hspace{3cm}   \\
\hspace{1cm} Michael {\it et al.} \cite{Michael}  \hspace{1cm}   & 2 224 579 (13) 
\hspace{3cm}   \\
\end{tabular}
\end{center}
\caption{\label{table1}The recently reported values of the deuteron
binding energy.  The deviations among the values show up at 10 eV order, 
which is just the same order as the Doppler effects contribute (see Eq. 
(\ref{dop})).  However notice that some of them have the error of several
electron volts order, which is smaller than the Doppler effects.}
\end{table}

\begin{table}
\begin{center}
\begin{tabular}{|c|c|c|c|}
(n,$\gamma$) reaction product nucleus (MeV) & $\gamma$-ray energy (MeV) &
Binding energy (MeV) & Doppler shift \\ \hline\hline
${}^3$ T & 6.250 316 & 6.257 268 (24) & 15.2 eV $\cos \theta$ \\
${}^{13}$ C & 4.949 319 & 4. 946 329 (24) & 2.8 eV $\cos \theta$ \\
${}^{14}$ C & 8.173 922 & 8.176 483 (40) & 4.2 eV $\cos \theta$ \\
${}^{15}$ N & 10.829 101 & 10.833 297 (38) & 5.3 eV $\cos \theta$ \\
\end{tabular}
\end{center}
\caption{\label{table3} These are other nucleus binding energies from
(n,$\gamma$) reactions in Ref. [5] and the possible Doppler shifts according
to the angle of the incident neutrons with respect to the direction of the
detected photons, where we assume that the kinetic energy of thermal neutrons
is 0.025 eV.}
\end{table}                                              
\end{document}